\documentclass[final,5p,times,twocolumn]{elsarticle}

\newcommand{\beq}{\begin{equation}}
\newcommand{\eeq}{\end{equation}}
\usepackage{amssymb}
\usepackage{amsbsy}   

\journal{Physica C}

\begin{document}

\begin{frontmatter}
 
\title{Explanation of the Meissner Effect and Prediction of a Spin Meissner Effect in Low and High $T_c$ Superconductors}


\author{J.E. Hirsch}

\address{Department of Physics, University of California, San Diego, La Jolla, CA 92093-0319}

\begin{abstract}
I argue that the conventional BCS-London theory of superconductivity does not explain the most fundamental property of superconductors, the Meissner effect: how is the  Meissner current
generated, and how is it able to defy Faraday's law? How is its mechanical angular momentum compensated?  I propose  that superconductivity
is impossible unless  the metal expels charge from its interior  towards the surface in the transition to superconductivity. As a consequence, superconductors in their ground state are predicted to possess a macroscopic electric field in their interior, as well as excess negative charge and a macroscopic spin current near the surface. The system is driven normal when the applied magnetic field is strong enough to bring the spin current to a stop. High temperature superconductivity occurs in systems that have too much negative charge.

\end{abstract}


\end{frontmatter}


When a metal is cooled into the superconducting state in the presence of an external magnetic field, a surface current spontaneously arises that generates a magnetic field in direction opposite to the external one, 
thus canceling the magnetic field in the interior of the superconductor (Meissner effect)\cite{meissner}. The conventional BCS theory of superconductivity\cite{bcs} describes the final state of this process, i.e. the superconducting state
with magnetic field excluded. However it does not explain how the Meissner current is generated, nor how the mechanical angular momentum carried by the Meissner current is compensated.

The equation of motion for an electron of charge $e$ and mass $m_e$  in the presence of electric and magnetic fields is
\beq
\frac{d\bold{v}}{dt}=\frac{e}{m_e}\bold{E}+\frac{e}{m_ec}\bold{v}\times\bold{B}
\eeq
Using
\beq
\frac{d\bold{v}}{dt}=\frac{\partial \bold{v}}{\partial t} + (\bold{v}\cdot \bold{\nabla})\bold{v}= \frac{\partial \bold{v}}{\partial t} +\bold{\nabla}(\frac{\bold{v}^2}{2})-\bold{v}\times(\bold{\nabla}\times\bold{v})
\eeq
and Faraday's law $\bold{\nabla} \times \bold{E}=-(1/c)\partial \bold{B}/\partial t$  it follows that
\beq
\frac{\partial \bold{w}}{\partial t} = \bold{\nabla}\times(\bold{w}\times\bold{v})
\eeq
for the `generalized vorticity'
\beq
\bold{w}=\bold{\nabla}\times\bold{v}+\frac{e}{m_ec}\bold{B} .
\eeq
Note that $\bold{w}$ is essentially the curl of the canonical momentum $\bold{p}=m_e \bold{v}+(e/c)\bold{A}$, with $\bold{A}$ the magnetic vector potential. 
In the normal state, at time $t=0$:
\beq
\bold{w}(\bold{r},t=0)=\frac{e}{m_ec}\bold{B}(t=0)\equiv \bold{w}_0
\eeq
independent of position $\bold{r}$. We set $\bold{\nabla}\times\bold{v}=0$ because in the normal state there is no net macroscopic charge flow.
Hence the canonical momentum $\bold{p}$ is nonzero throughout the interior of the superconductor in the initial state. 

In the superconducting state, the superfluid velocity $\bold{v}$ obeys the London equation\cite{bcs}
\beq
\bold{\nabla}\times\bold{v}=-\frac{e}{m_ec}\bold{B}
\eeq
Together with Ampere's law, Eq. (6)  implies that the Meissner current flows within a London penetration depth
\beq
\lambda_L=(\frac{m_e c^2}{4\pi n_s e^2})^{1/2}
\eeq
of the surface, with $n_s$ the superfluid carrier density. Therefore, according to Eqs. (4) and (6)
\beq
\bold{w}(\bold{r},t=\infty)=0
\eeq
everywhere in the superconducting body. 
Equivalently,  the canonical momentum $\bold{p}=0$ throughout the interior of the (simply connected)  superconductor.
BCS theory predicts that the state of lowest energy for the superconductor in the presence of a magnetic field (that is lower than the
critical field) has $\bold{p}=0$, or equivalently $\bold{w}=0$. However, 
neither London theory nor BCS theory address the question of $how$ $\bold{w}$ changes from its initial value 
$\bold{w}_0$  to its final value $\bold{w}=0$.

In a cylindrical geometry, assuming azimuthal symmetry as well as translational symmetry along the cylinder axis ($z$) direction (infinitely long cylinder)
\beq
\bold{w}(\bold{r},t)=w(r,t)\hat{z}
\eeq
and Eq. (3) takes the form
\beq
\frac{\partial w}{\partial t} =-\frac{1}{r}\frac{\partial}{\partial r}(rwv_r)
\eeq
with $r$ the radius in cylindrical coordinates. Eq. (10)  implies that  {\it  $ w $ can only change if there is radial flow of charge} ($v_r \neq 0$). Moreover,  for $w$ to evolve towards
its final value $0$ requires $v_r>0$, i.e. a radial $outflow$ of electrons. BCS theory does not predict a radial outflow of electrons in the transition to
superconductivity, hence within Eq. (10) it predicts that $w$ does not change with time. 
Consequently, BCS theory claims that in the transition to superconductivity the conduction electrons ignore the fundamental
equation of motion Eq. (1) and somehow magically evolve $\bold{w}$  from its initial non-zero value throughout the interior of the superconductor to zero everywhere.
Perhaps through discrete jumps in an undetectable fractal hyperspace?
We argue that this is unphysical and impossible. Instead, the theory of hole superconductivity predicts expulsion of negative charge from the interior of the
superconductor\cite{expulsion}, hence $v_r\neq 0$, hence provides an explanation for how $\bold{w}$ is driven to zero.

Eq. (10) can be written as
\beq
\frac{\partial w}{\partial t}=-\bold{\nabla}\cdot(w \bold{v})
\eeq
which has the form of a continuity equation. It implies that the only way that $w$ can change in a region enclosed by a surface is by {\it flowing out through
the surface}, carried by the fluid flow moving with velocity $\bold{v}$. We conclude that in order for  $\bold{w}$ to evolve from $\bold{w}_0$ to zero,
superfluid electrons have to flow out from the cylinder through its surface, thereby carrying $\bold{w}$ outside the superconductor.
This outflow of electrons when the metal is cooled below $T_c$ and expels a magnetic field from its interior should
be observable through   optical or X-ray diffraction techniques with grazing incidence on a clean metal surface.

If there is a radial outflow of electrons when a system goes superconducting, the final state will not have a homogeneous charge distribution, hence there will be an electric field in the interior
of the superconductor. The conventional London equations cannot describe such situation because  the `first London equation'
$\partial \bold{J}/ \partial t =(ne^2/m_e) \bold{E}$
implies that the superfluid current $\bold{J}$ diverges in the presence of an electric field. Instead, the modified London equations proposed by
the author\cite{electrodynamics} and by London\cite{londonold} allow for the presence of a static electric field in the interior of superconductors. These equations are derived
in a manner similar to the conventional derivation of the London equations except that the Lorentz rather than the London gauge is assumed to
be valid in the second London equation
$\bold{J}=-(ne^2/m_e c)\bold{A}$, 
resulting in the constitutive relation\cite{electrodynamics}
\beq
\phi-\phi_0=-4\pi\lambda_L^2(\rho-\rho_0)
\eeq
relating the electric potential $\phi$ to the charge density $\rho$. The parameter $\rho_0>0$ is an integration constant giving the uniform positive
charge density in the interior of the superconductor resulting from the expulsion of negative charge, and $\phi_0(\bold{r})$ is the corresponding potential. For a cylindrical geometry, the electrostatic field
in the interior far from the surface is given by\cite{electrospin}
$ \bold{E}(\bold{r})=-\bold{\nabla}\phi(\bold{r})=E_m \bold{r}/R$,  
with
\beq
E_m=-\frac{\hbar c}{4e\lambda_L^2}
\eeq
and $R$ the radius of the cylinder.

The expulsion of negative charge from the interior of the superconductor gives rise to a macroscopic spin current flowing within a London penetration depth of the
surface of the superconductor\cite{sm}, with carrier velocity given by
\beq
\bold{v}_\sigma ^0=-\frac{\hbar}{4m_e \lambda_L} \pmb{\sigma} \times \bold{n}
\eeq
with $\pmb{\sigma}$ the carrier's spin and $\bold{n}$ the unit vector in direction normal to the surface. The associated kinetic energy satisfies
\beq
\frac{1}{2} m_e (v_\sigma^0)^2 n_s=\frac{E_m^2}{8\pi}
\eeq
which is analogous to the equation relating the superfluid charge velocity to the magnetic energy density\cite{bcs}.

In the presence of an external magnetic field $\bold{B}$ the carrier velocity is
$\bold{v}_\sigma=\bold{v}_\sigma^0-(e/m_ec)\bold{A}$,
and the magnetic field that stops one of the components of the spin current (sets $v_\sigma=0$) has magnitude
\beq
B_s=\frac{m_e c}{e\lambda_L} v_\sigma^0=-\frac{\hbar c }{4e\lambda_L^2}=E_m
\eeq
and is similar to $H_{c1}$, the lower critical field in the conventional theory\cite{bcs}. The external magnetic field induces a non-zero spin density near the surface\cite{electrospin}
($\uparrow$ is the direction of $\bold{B}$)
\beq
n_\uparrow - n_\downarrow = \frac{1}{4\pi e \lambda_L}B
\eeq
which will give rise to a non-zero Knight shift at zero temperature.

The driving force for charge expulsion is kinetic energy lowering associated with expansion of the electronic orbits from radius $k_F^{-1}$ to radius
$2\lambda_L$\cite{sm}. This is the origin of the `electromotive force' that overcomes the Faraday counter-emf arising when the magnetic
field is expelled, as well as the electrostatic potential energy rise created by the charge expulsion. 
This radial  `quantum force' (or `quantum pressure') associated with expansion of the de Broglie wavelength  is very different from an ad-hoc
azimuthal `quantum force' postulated by Nikulov in an attempt to explain the Little-Parks experiment and the Meissner effect\cite{nikulov}. How the angular momentum in
the Meissner current is compensated is discussed in Ref. \cite{sm}.

Metals will have strongest tendency to expel negative charge when their conduction band has many electrons (is almost full, i.e. holes) and when the conducting substructures
(eg planes) have excess negative charge, as occurs in the cuprates, arsenides and $MgB_2$. Those are the ideal conditions for high temperature superconductivity,
but the essential physics is proposed to be the same for all superconductors, whether $T_c$ is high or low.




\end{document}